\documentclass{pasj00}
\draft
\usepackage{lscape}
\usepackage{comment} 

\newcommand{\msun}{\ensuremath{M_{\odot}}}
\newcommand{\rhob}{\ensuremath{\rho_{\rm B}}}
\newcommand{\esym}{\ensuremath{E_{\rm sym}}}
\newcommand{\yc}{\ensuremath{Y_{\rm C}}}
\newcommand{\nsmax}{\ensuremath{\rm M_{\rm NS}^{\rm Max}}}
\newcommand{\nucm}[2]{\ensuremath{{}^{#1}{\rm #2}}}
\newcommand{\hyp}{\ensuremath{\, \mathchar`- \,}}

\begin{document}
\SetRunningHead{C. Ishizuka et al.}{The EOSDB}

\title{EOSDB: The Database for Nuclear EoS \thanks{http://aspht1.ph.noda.tus.ac.jp/eos/}}

\author{Chikako \textsc{Ishizuka} %
}
\affil{Research Institute for Science and Technology, Tokyo University of Science, Yamazaki 2641, Noda, Chiba 278-8510, Japan}
\email{chikako@rs.tus.ac.jp}

\author{Takuma \textsc{Suda}}
\affil{Research Center for the Early Universe, The University of Tokyo, Hongo 7-3-1, Bunkyo-ku, Tokyo 113-0033, Japan}\email{suda@resceu.s.u-tokyo.ac.jp}

\author{Hideyuki {\sc Suzuki}}
\affil{Faculty of Science and Technology, Tokyo University of Science, Yamazaki 2641, Noda, Chiba 278-8510, Japan}\email{suzukih@rs.tus.ac.jp}

\author{Akira \textsc{Ohnishi}}
\affil{Yukawa Institute for Theoretical Physics, Kyoto University, Kitashirakawa Oiwakecho, Sakyo-ku, Kyoto 606-8502, Japan}\email{ohnishi@yukawa.kyoto-u.ac.jp}

\author{Kohsuke \textsc{Sumiyoshi}}
\affil{Numazu College of Technology, Ooka 3600, Numazu, Shizuoka 410-8501, Japan}\email{sumi@numazu-ct.ac.jp}
\and
\author{Hiroshi \textsc{Toki}}
\affil{Research Center for Nuclear Physics, Osaka University, Mihogaoka 10-1, Ibaraki, Osaka 567-0047, Japan}\email{toki@rcnp.osaka-u.ac.jp}

%


\maketitle

\begin{abstract}
Nuclear equation of state (EoS) plays an important role in understanding
the formation of compact objects such as neutron stars and black holes.
The true nature of the EoS 
has been a matter of debate at any density range not only in the nuclear physics
but also in the astronomy and astrophysics.
We have constructed a database of EoSs by compiling data from the literature.
Our database contains the basic properties of the nuclear EoS of symmetric nuclear matter
and of pure neutron matter.
It also includes the detailed information about the theoretical models, for example the adopted 
methods and assumptions in individual models.
The novelty of the database is to consider new experimental probes such as
the symmetry energy, its slope with respect to the baryon density, and the incompressibility,
which enables users to check their model dependences.
We demonstrate the performance of the EOSDB through the examinations of the model
dependence among different nuclear EoSs.
It is reveled that some theoretical EoSs, which is commonly used in astrophysics, do not
satisfactorily agree with the experimental constraints.
\end{abstract}

\section{Introduction}\label{sec:intro} 
Nuclear equation of state (EoS)
describes the properties of dense nuclear matter whose density typical ranges $10^{9-15}$ g/cm$^{3}$.
It plays an important role both in nuclear physics and astrophysics,
because the EoS of dense matter is directly related to heavy nuclei as well as dense matter in 
compact objects such as neutron star and black holes after supernova explosions
(SNe).
For example, simulations of neutron star mergers and/or SNe have
been performed to constrain nuclear models \citep{Hotokezaka2013,Bauswein2012}. 
Furthermore, in an effort to connect nuclear EoSs and the hydrodynamical simulations of SNe,
\citet{Typel2013} developed the database of EoSs for SNe (CompOSE by the CompStar team) at
finite temperature systems. It provides EoS tables with nuclear statistical equilibrium (NSE)
approximation for the inhomogeneous phase below the nuclear saturation density.

From the observational point of view,
the X-ray observations of neutron stars can constrain nuclear EoSs
through the determination of masses and radii using the Tolman-Oppenheimer-Volkoff equation,
although these quantities are influenced by the model assumptions for neutron star atmospheres. 
Once the mass and radius of a neutron star are determined observationally,
the EoSs can be well constrained by the possible structure properties of neutron stars
inferred from its total mass and radius.
Therefore, the discoveries of massive neutron stars with $\sim 2 \msun$ 
(PRS J1614-2230 with 1.97 $\pm$ 0.04\msun~\citep{Demorest2010} 
and PRS J0348+0432 with 2.01 $\pm$ 0.04 \msun~\citep{Antoniadis2013})
cast a doubt on the the existing EoSs derived from nuclear physics.
Since the central densities become high enough,
exotic constituents and exotic states such as hyperons (baryons with strange quarks),
meson condensation, and quarks, are expected to appear, but 
most of the proposed EoSs with exotic constituents cannot sustain massive neutron stars
~\citep{Lattimer-Prakash2010}.

Several ideas have been proposed to retain the consistency between astronomical observations and
laboratory experiments.
\citet{Miyatsu2012} consider the inter-baryon interactions to suppress the appearance of hyperons.
Adjusting hyperon-nucleon or hyperon-hyperon interactions is another idea to make the EoSs stiff.
\citet{Weissenborn2012} proposed an EoS that is stiff enough to support massive neutron stars using 
meson octet, singlet coupling contents and meson-hyperon coupling strengths as fitting parameters.
\citet{Sulaksono2012} improved their idea to satisfy nuclear experimental results
by adjusting the strengths of these interactions, which are experimentally unknown at present.
\citet{Masuda2013} assumed crossover transition from nuclear phase to quark phase
to support $2 \msun$ neutron stars.
On the other hand, \citet{Tsubakihara2013} argued that the three body interactions are to be investigated to realize stiffer EoSs.
More investigations are ongoing with ab initio calculations of nuclear EoSs~\citep{Takano2010,Aoki2012}.
The magnetic field may also change the effective EoS 
because single pulsars have $10^3$ times stronger magnetic field ($10^{14 - 15}$ G) than typical neutron stars do.
\citet{Potekin2012} and \citet{Myung-Ki2013} insist that
such strong field can be responsible for stiff EoSs even without any other additional interactions.
Thus, there exists hundreds of published EoSs
from nuclear physics community with state-of-the-art input physics taken into account.

Contrary to the intensive explorations of the EoS by nuclear physics community,
EoSs adopted in astrophysical context are very limited.
Thanks to the many efforts to provide EoSs in astrophysics, improved nuclear EoSs have been
available, in addition to the present standard EoSs, 
e.g., polytropic EoSs, non-relativistic Lattimer-Swesty's EoS \citep{LS} which advanced
the first nuclear EoS for astronomical use~\citep{Hillebrandt1984},
and relativistic Shen's EoS \citep{Shen1998}.
Crucial aspect for the astrophysics application of nuclear EoSs has been
the incompressibility $K = 9\rho_0^2\partial^2E/\partial\rho_B^2|_{\rho_B = \rho_0}$
at the saturation density $\rho_B = \rho_0 \simeq 0.16$ fm$^{-3}$, which is related to recoil of compressed materials.
The incompressibility of nuclei could be a key to determine the maximum mass of neutron stars.
It has been studied by using nuclear compression modes, such as Iso-Scalar Giant Monopole Resonance (ISGMR) 
and Iso-Scalar Giant Dipole Resonance (ISGDR).
The recommended value of the nuclear matter incompressibility is about $230 \hyp 270$ MeV estimated from 
\nucm{208}{Pb} and \nucm{144}{Sm} data
using different types of interactions \citep{Colo-Giai}.
At present, these properties of symmetric nuclear matter have been well determined 
both in nuclear theories and experiments.
It is well known that non-relativistic models give smaller incompressibility,
while relativistic models give larger incompressibility.
This model dependence comes from the treatment of nuclear surface,
which is always involved in the experimental data.
\citet{LS} use three choices of the values
for incompressibility that gives a reasonable constraint on nuclear EoSs.
\citet{Shen1998}, on the other hand, provided an EoS for the first time in tabular format
with relativity taken into consideration to be applicable in astronomical condition where relativity becomes important.
This is also the first EoS table for astrophysics applications considering experimental information on neutron-rich and heavy nuclei 
that correspond to charge asymmetry and matter-like property 
of nuclear many-body system.

In order for a EoS to be realistic, 
there are three requirements to be fulfilled; the saturation point,
the incompressibility, and the symmetry energy.
The symmetric nuclear matter has its minimum energy (per nucleon) $E = E_0 \simeq -16$ MeV 
at the saturation density.
This is the so-called ``nuclear saturation property''.
In addition to these criteria, 
symmetry energy and the slope of the symmetry energy with respect to baryon density, have drawn
much attention in the last decade.
The symmetry energy is also a dominant component of the bulk nuclear property,
and has a great impact on the understanding of pure neutron matter.
This is because these quantities can indirectly constrain the property of 
pure neutron matter that is difficult to know due to the lack of 
direct experimental approaches.
It is to be noted that the symmetry energy can be considered the energy difference
between symmetric nuclear matter and pure neutron matter as its $0 \hyp$th order approximation.
Thanks to the efforts to estimate the symmetry energy, there are several plans of
experiments to constrain the EoSs at very low densities.
The symmetry energies are expected to be determined experimentally as a function of baryon densities
for $(1/3 \hyp 1) \rho_0$ at MSU, $(1 \hyp 2) \rho_0$ at RIKEN, and $(2 \hyp 3) \rho_0$
at GSI by using the mass formula, 
isobaric analogue state (IAS), heavy-ion collisions(HIC), and 
neutron skin thickness.
The evaluated value of the symmetry energy $E_{\rm sym}$ and its slope $L$ 
is $31 \pm 3$ MeV and  $54 \pm 13$ MeV, respectively \citep{Chen2010}.
As for the determination of pure neutron matter EoSs at extremely low densities, 
they can be directly measured by cold atom of dilute Fermi gas.

These experimental constraints on nuclear properties generally include model-dependence
in their analysis procedures. Especially the symmetry energy is very difficult
to measure directly with the current experimental techniques,
because of the mutual dependence between experimental analyses and nuclear models at high densities.
 
To overcome the difficulty regarding the constraint on the EoSs in an appropriate way,
we have constructed a database for nuclear EoSs (EOSDB)
to assemble as many data dealing with all the four criteria discussed above as possible from the literature.
This enables us to integrate the pieces of information about the constraints on nuclear EoSs available in the literature because
there are few papers that satisfy all these criteria
due to various concerns regarding the properties of nuclear matter.
In assembling the data, we pay much attention to the model dependences
ascribable to the adopted EoSs.
In particular, it is difficult to remove the model dependences caused by the symmetry energy
since it can only be derived from raw experimental data using theoretical models.
The new database system will help to check the properties of each data
through the comparisons of different EoSs in a unified scheme.
The EOSDB provides EoSs together 
with the nuclear saturation properties, the symmetry energy properties
and the information related to the mass and radius of neutron stars.
The database is also designed to analyze model dependence
by compiling theoretical models and assumptions adopted in
each EoS.

The paper is organized as follows:
The details of the EOSDB are described in \S 2.
\S 3 devotes the description about the usage.
The example of a model analysis using the EOSDB is given in \S 4.
Summary and discussions follow in \S 5.

\section{Characteristics of the EOSDB}\label{sec:chardb}
The basic structure of the EOSDB is common with the SAGA database \citep{Suda2008,Suda2011}
that is a database for observed metal-poor stars.
The EOSDB is operated by MySQL and CGI.
The retrieval and data plotting systems are constructed
with Perl and JavaScript. We prepared libraries for compilation
of the EoSs; the list of major journals, 
basic physics constants that are used in the calculation,
classifications for constituents of dense matter, methods,
thermodynamical variables, and
physical quantities on symmetric energies.
A unit record is defined by the data of interaction available in individual papers.
Each record has compiled data according to the format defined by the library.
The most important quantities in the database are
those related to the basic EoS properties such as thermodynamical quantities,
the symmetry energy $E_{\rm sym}$, its slope $L$, and the incompressibility $K$
as a function of baryon densities for various models.

The selection criteria of papers are as follows with the decreasing order of priority: 
\begin{itemize}
\item Articles containing data distributed to the public
\item Articles often cited as a standard EoS (e.g., Lattimer \& Swesty, Shen, Akmal-Pandharipande-Ravenhall)
\item Articles containing constraints on the EoSs
\end{itemize}
Most of the compiled data
are derived from theoretical models, although some experimental and observational constraints 
are also included. 
Table~\ref{tab:compiled_data} gives the list of compiled papers,
selected from hundreds of candidate papers dealing with EoSs published in the last decade.
It is to be noted that 
the number of candidate papers has increased drastically after the discovery of the massive neutron stars.

In the current version of the database, all of the records are based on the models of $T = 0$ MeV for symmetric nuclear matter 
and/or pure neutron matter.
This is because we can focus on the most basic features of hadronic matter, and see the difference among models and the behavior of the models.
If the basic parameters of the EoSs cannot be derived from theoretical models at the exact zero temperature,
we adopt those for a finite temperature, trying to use as low temperature as possible. 

\begin{longtable}{lllll}
\caption{List of Compiled data in the EOSDB}\label{tab:compiled_data}
     \hline   
     Data ID                & Code            & Constituents           &             & \\
                            & (Reference)     &                        &             & \\
      $\rho_0$[fm$^{-3}$] \ & \ $E_0$ [MeV] \ &  $E_{\rm sym}$ [MeV] \ & $L$ [MeV] \ & \ $K$ [MeV] \\
      \hline
\endhead
\hline
       \multicolumn{5}{l}{\hbox to 0pt{\parbox{85mm}{\footnotesize
	    \par\noindent
	    \footnotemark[$*$] The corrected values when we adjust the binding energy to -16 [MeV]. 
	    \par\noindent
	    \footnotemark[$\dagger$] Experimental constraint on EoSs.
	    \par\noindent
	    \footnotemark[$\ddagger$] Observational constrain on EoSs.
	     \par\noindent
	    \footnotemark[$\S$] Data ready to compiled as of 24, Sep., 2013.	    
	 }}}
\endfoot      
E0001 & GShenPRC2011\_FSUgold2.1& $n, p, \alpha, A$ & & \\
 &  ~\citep{GShen} & & & \\
0.148 & -16.30 & 32.59 & 60.5 & 230 \\
      \hline
E0002 & HShenNPA1998 & $n, p, \alpha, A$ & & \\
 & ~\citep{Shen1998} & & & \\
0.145 & -16.3 & 36.9 & 110.8 & 281 \\
      \hline
E0003 & HShenAPJS2011\_N & $n, p, \alpha, A$  & & \\
 & ~\citep{Shen2011} & & & \\
0.145 & -16.3 & 36.9 & 110.8 & 281 \\
      \hline
E0004 & HShenAJPS2011\_Y & $n, p, \alpha, A, \Lambda$  & & \\
 & ~\citep{Shen2011} & & & \\
0.145 & -16.3 & 36.9 & 110.8 & 281 \\
      \hline
E0005 & LatttimerNPA1991\_LS180 &  $n, p, \alpha, A$  & & \\
 & ~\citep{LS} & & & \\
0.155 & -16 & 29.3 & 74 & 180 \\
      \hline
E0006 & LattimerNPA1991\_LS220 &  $n, p, \alpha, A$ & & \\
 & ~\citep{LS} & & & \\
0.155 & -16 & 29.3 & 74 & 220 \\
      \hline
E0007 & LattimerNPA1991\_LS375 &  $n, p, \alpha, A$ & & \\
 & ~\citep{LS} & & & \\
0.155 & -16 & 29.3 & 74 & 375 \\
      \hline
E0008 & HempelNPA2010\_TMA  & $n, p, \alpha, A$ & & \\
 & ~\citep{Hempel2010} & & & \\
0.147 & -16.03 & 30.66 & 90.14 & 318 \\
      \hline
E0009 & MiyatsuPLB2012 & $n, p, \Lambda, \Sigma^{0, \pm}, \Xi^\pm$ & & \\
 & ~\citep{Miyatsu2012} & & & \\
0.15 & -15.7 & 32.5 & 88.7 & 280 \\
      \hline
E0010 & KanzawaPTP2009& $n, p$ & & \\
 & ~\citep{Kanzawa}  & & & \\
0.16 & -16.09 & 30.0 & --- & 250 \\
      \hline
E0011 & FurusawaApJ2011 & $n, p, \alpha, A$ & & \\ 
 & ~\citep{Furusawa2011} & & & \\
0.145 & -16.3 & 36.9 & 110.8 & 281 \\ 
      \hline
E0012 & IshizukaJPG2008\_SR30 & $n, p, \alpha, A, \Lambda, \Sigma^{0, \pm}, \Xi^\pm$ & & \\ 
 & ~\citep{IOS2008} & & & \\
0.145 & -16.3 & 36.9 & 110.8 & 281 \\
      \hline
E0013 & HillebrandtAA1984& $n, p, A$ & & \\
 & ~\citep{Hillebrandt1984}  & & & \\
0.155 & -16.0 & 32.9 & --- & 263 \\
      \hline
E0014 & TimmesAPJS1999 & Helmholtz type EoS & & \\
 & ~\citep{Timmes}  & & & \\
--- & --- & --- & --- & ---\\
      \hline
E0015 & NewtonJPC2006 & $n, p$ & & \\
 & ~\citep{Newton} & & & \\
0.16 & -15.78 & 30.03 & 45.78 & 216.7 \\
      \hline
E0016 & AkmalPRC1998\_AV18 & $n, p$ & & \\ 
 & ~\citep{APR1998} & & & \\
0.16 & -14.59 & 32.60$^\dagger$ & 57.6 & 266.0\footnotemark[$*$] \\
      \hline             
E0017 & AkmalPRC1998\_AV18\_3BF & $n, p$ & & \\
 & ~\citep{APR1998} & & & \\
0.16 & -11.85 & 32.60$^\dagger$ & 57.6 & 266.0\footnotemark[$*$]\\
      \hline
E0018 & AkmalPRC1998\_AV18\_Boost & $n, p$ & & \\
 & ~\citep{APR1998}  & & & \\
0.16 & -12.54 & 32.60$^\dagger$ & 57.6 & 266.0\footnotemark[$*$] \\
      \hline
E0019 & AkmalPRC1998\_AV18\_3BF\_Boost & $n, p$  & & \\
 & ~\citep{APR1998}  & & & \\
0.16 & -12.16 & 32.60$^\dagger$ & 57.6 & 266.0\footnotemark[$*$] \\
 \hline   
E0020 & ZuoNPA2002 & $n, p$ & & \\
 & ~\citep{Zuo} & & & \\
0.198 & -15.08 & --- & --- & 207\\
      \hline
E0021 & GrossNPA1999 & $n, p$ & & \\ 
 & ~\citep{Gross} & & & \\
0.185 & -16.15 & 34.36 (31.6@0.16 fm$^{-3}$) & ---& 230 \\
      \hline
E0022 & vanDalenNPA2004 & $n, p$ & & \\ 
 & ~\citep{Dalen2004}  & & &  \\
0.185 & -16.15 & 34.36(31.6@0.16 fm$^{-3}$ & --- &230 \\
      \hline
E0023 & TypelNPA1999 & $n, p$ & & \\ 
 & ~\citep{Typel1999} & & & \\
0.153 & -16.247 & 33.39 & --- & 240\\
      \hline
E0024 & NiksicPRC2002\footnotemark[$\dagger$] & $n, p$ & & \\
 & ~\citep{Niksic2002} & & & \\
0.152 & -16.20 & 33.1 & 55 & 244.5 \\ 
	\hline
E0025\footnotemark[$\S$] & SteinerPRC2005 & $n, p$ & & \\
 & ~\citep{Steiner2005} & & & \\
0.16 & -16 & 31.6 & 107.4 & 211 \\ 
      \hline      
E0026\footnotemark[$\S\dagger$] & TsangPRC2012\footnotemark[$\dagger$] & --- & & \\
 & ~\citep{Tsang2012} & & & \\ 
0.16 & --- & 30 $\le E_{sym} \le$34.4 & 45 $\le L \le$ 110 & --- \\        
      \hline
E0027\footnotemark[$\S\dagger$]  & DanielewiczRSEPSN2002\footnotemark[$\dagger$] & --- & & \\
 & ~\citep{Danielewicz2002} & & & \\
--- & --- & --- & --- & 300 \\  
      \hline
E0028\footnotemark[$\S$] & SteinerPRL2012\footnotemark[$\ddagger$] & only $n$ & & \\
   &  \citep{Steiner2012} & & & \\
--- & -16 & 32 $\le E_{sym} \le$ 34& 43 $\le L \le$ 52 & --- \\ 
      \hline
E0029\footnotemark[$\S$] & OnoPTPS2002 & $n, p$ & & \\
 & ~\citep{Ono2002} & & & \\
--- & -16.32 & 30.8 & --- & 228.0 \\       
      \hline
E0030\footnotemark[$\S$] & FriedmanNPA1981 & $n, p$ & & \\
 & ~\citep{Friedman1981} & & & \\
0.16 & --- & --- & --- & 240 \\        
      \hline
E0031\footnotemark[$\S$] & CarlsonPRC2003 & --- & & \\
 & ~\citep{Carlson2003}  & & & \\
--- & --- & --- & --- & --- \\       
      \hline
E0032\footnotemark[$\S$] & GezerlisPRC2010 & only $n$ & & \\
 & ~\citep{Gezerlis2010} & & & \\
--- & --- & --- & --- & --- \\        
      \hline
E0033\footnotemark[$\S$] &GandolfiPRC2009 & only $n$ & & \\
 & ~\citep{Gandolfi09PNM} & & & \\
--- & --- & --- & --- & --- \\        
      \hline
E0034\footnotemark[$\S$] & GandolfiPRL2007 & only $n$ & & \\
 & ~\citep{Gandolfi07SNM} & & & \\
--- & --- & --- & --- & --- \\     
      \hline  
E0035\footnotemark[$\S$] & SchwenkPRL2005 & only $n$ & & \\
 & ~\citep{Schwenk2005} & & & \\
0.16 & --- & --- & --- & --- \\    
      \hline  
E0036 & BotvinaNPA2010 & n, p, $\alpha$, A & & \\
 & ~\citep{Nihal2012} & & & \\
0.145 & -16.3 & 36.9 & 110.8 & 281 \\ 
      \hline
\end{longtable}

As of Aug. 2014,
36 data sets have been compiled (see Table~\ref{tab:compiled_data}).  
The data ID and the reference ID in the first and second column, respectively, specifies the record in the database.
The IDs can be found on the web site in using the data search and plot system (See \S~\ref{sec:system}).
The other columns in the tables describe the quantities related to the saturation properties of the data.
Constituents denote the components considered in the data.
The quantities in each bottom line, $\rho_0$ [fm$^{-3}$], $E_0$ [MeV] , 
$E_{\rm sym}$ [MeV], $L$ [MeV] , $K$  [MeV]
are saturation density, binding energy, symmetry energy, symmetry energy slope to baryon density at $\rho_0$,
and incompressibility, respectively.
The details about these quantities are described in the following subsections.
Energy except for the saturation point, and the other quantities such as pressure and entropy,
can be downloaded from the online database.

All the data compiled in the database are also stored in text format and
are accessible through the web site so that users can inspect individual data
in more detail.
The text data include the following information.
\begin{enumerate}
\item Bibliographic information 
\item Instructions on how to handle the tabulated data if exists
\item Physics constants used in each EoS
\item Assumed constituents and conditions
\item Theoretical/Experimental/Observational methods to derive each EoS and its strong/weak points and comments.
\item Saturation density
\item Saturation energy
\item Symmetry Energy properties
\item Maximum cold neutron star mass (if calculated data exist) 
\item Source of data (if tabulated data or numerical codes are distributed or not.)
\end{enumerate}
The EOSDB web site also provides the link to the original papers in which full information should be available.

The data of nuclear EoSs are taken from open EoS tables 
and/or compiled data using a software named GSYS
\footnote{distributed by the JCPRG at http://www.jcprg.org/gsys/gsys-e.html}
to read viewgraphs.
Most of data in the EOSDB has been scanned from viewgraphs in the papers using the GSYS at present.
These data possibly include systematic errors because the work relies on manual operation in using the software.

Table~\ref{tab:quantity} presents physical quantities registered in the database.
In the following, we give their details along with our categories and items.
As explained in the previous section, these quantities are essential to characterize and constrain nuclear models.

\begin{table}
  \caption{Data compiled in the EOSDB}
  \label{tab:quantity}
  \begin{center}
    \begin{tabular}{*{3}{l}}
      \hline
      Data table category & Item & Note \\
      \hline
      Bibliography & Title & \\
                   & Authors & \\
                   & Reference  & \\
      \hline
      Attribute & Theory & For pure theoretical arguments\\
                    & Pure Experiment & Experimental constraints on EoSs\\
                    & Analysis & Theoretical analysis of experimental results\\             
      \hline
      Constituents & N, Y, $\alpha$, A, Q, L & particles or nuclei\\
      \hline
      Method & Model &  Theoretical framework \\
                   & Approximation & \\
      \hline             
      Physics Constants & $\hbar, c$, amu, etc. & \\             
      \hline
      EoS for SNM\footnotemark[$*$] & \rhob &  Baryon density\\
                 &  $Y_{\rm C}$ & Charge Ratio $Y_{\rm C} = 0.5$  \\
                 &  $E/B$ & Energy per baryon\\
                 &  $P$ & Pressure \\
                 &  $S$  & Entropy\\
      \hline                
      EoS  for PNM\footnotemark[$\dagger$] & \rhob &  Baryon density\\
                 & $Y_{\rm C}$ & Charge Ratio $Y_{\rm C}=0.0$  \\
                 &  $E/B$ & Energy per baryon\\
                 &  $P$ & Pressure \\
                 &  $S$  & Entropy\\      
      \hline
      Symmetry Energy   & \esym & Symmetry energy  \\
                    & $L$ & \esym\ slope to baryon density  \\
                    & $K$ & Incompressibility \\                            
      \hline
      \multicolumn{3}{@{}l@{}}{\hbox to 0pt{\parbox{85mm}{\footnotesize
          \par\noindent
          \footnotemark[$*$] symmetric nuclear matter
          \par\noindent
          \footnotemark[$\dagger$] pure neutron matter
      }}}
    \end{tabular}
  \end{center}                                                                                                  
\end{table}

\subsection{Bibliography and Attribute}

The identifier of the compiled paper is one of the primary key of the database.
We include bibliographic data in the database and label them specific ID (Data ID) as
shown in the first top columns in Table~\ref{tab:compiled_data}.
We also give identifiers for the data in individual papers (Reference ID).
The reference ID is given by the following format;
[Surname of the first author][Journal code][Year]\_[Comment].
The comment in the ID is added only if
two or more EoSs can be compiled from a single paper.

Compiled data are classified according to which approach is made in the paper
to deduce a constraint on EoSs;
theory, experiment, or both.
The observational determinations on mass and radius of neutron stars are classified as
experimental approach in our database.
 
\subsection{Constituents, Methods, Physics Constants}
Constituents and methods adopted in the papers are helpful when users try 
to reproduce the original data by themselves. 
The constituents of nuclear matter that we registered are given in Table~\ref{tab:quantity}.
The symbols $N, Y, \alpha, A, Q$ and $L$ correspond to
nucleons, hyperons, $\alpha$-particles, nuclei quarks, and leptons, respectively.
The other exotic particles can be added to the library if needed.
Note that the $L$ in the list of constituents is different from the symmetry energy slope $L$.
They are treated as different quantities in the database.
The constituents in EoSs are important information in the database since they directly affect their energy and/or pressure.
Users should check the components of a system when they compare different sets of EoSs. 

The database contains information about the theoretical frameworks, approximations and assumptions 
adopted in the papers that are selected from the data in the library.
It is useful in surveying the model dependence of each physical quantity
as demonstrated in \S\ref{sec:examples}. 
We selected around 40 representative major theoretical frameworks, models, and approximations 
published in the last decade and registered them as method codes in the library.
If two or more models and approximations are used
(for example, the relativistic mean field and random phase approximation)
in one record, they are both compiled (``RMF'' and ``RPA'' should be used in this case).
In the future updates, users will be able to use the methods as a query option in search and plot system.

\subsection{EoS for $Y_{\rm C} = 0$ and $0.5$}\label{sec:eos} 
In the EOSDB, thermodynamical quantities for the charge ratio of $0.0$ and $0.5$ are compiled
as a function of baryon densities.
The charge ratio is defined as follows,
\begin{equation}
Y_{\rm C} = (\displaystyle{\sum_{i=n, p,...}}Q_i\rho_i
+ \displaystyle{\sum_{j=e^\pm, \mu^\pm}}Q_j\rho_j)/\rhob ,
\end{equation}
where $Q_i$ and $\rho_i$ is the charge and number density of particle $i$, respectively.

For $\yc = 0.0$, the database for pure neutron matter is expected to play
an essential role in imposing strong constraints on nuclear EoSs. 
However, the direct determinations of EoSs for pure neutron matter is extremely difficult
except for low deinsities where $\yc = 0.0$ can be reproduced in laboratory environment.
From the theoretical point of view, the predictions for nuclear EoSs for pure neutron matter
by different groups do not agree with each other, even for the same theoretical frameworks.
This has been a controversy among theorists and brought about a motivation to
compare different EoSs.
It is to be noted that the EOSDB has two cases for $\yc = 0.0$.
One is a pure neutron matter consisting of neutrons, and another is a neutron star matter
consisting of neutrons and a small number of protons and electrons.
It is not meaningful to compare EoSs with different composition even if the data are provided for $\yc = 0.0$.

If leptons are included in a system, the users should consider the contribution of leptons and need special care
when comparing with other EoSs.
Most of published data related to neutron stars contain leptons.
The reason we include leptons in the definition of $Y_{\rm C}$ is
that some theoretical models do not contain any published data 
without leptons, which is the case in typical neutron star calculations.
Whether or not leptons are included in the dataset is described
in the text data that the EOSDB web site provides.
Users can choose data without leptons only.

The dataset with $\yc = 0.5$ needs a special attention when they are used.
As discussed in \S\ref{sec:intro}, we can
investigate nuclear saturation properties from the EoS of symmetric nuclear matter.
For $Y_{\rm C} = 0.5$, we assume symmetric nuclear matter which consists of the equal 
number of neutrons and protons, or a charge symmetric system which includes hyperons 
(the other members of baryon octet). 
In the most of the datasets for $Y_{\rm C} = 0.5$, available EoSs are basically for symmetric nuclear matter.  
We compile all the energy data with $\yc = 0.5$
that provides nuclear saturation properties.
This enables to evaluate the validity of compiled EoSs by
verifying that the compiled energy satisfies
the typical value of the saturation energy of $-16$ MeV 
at the saturation density of 0.16 fm$^{-3}$. 
This check will be important for EoSs derived from theoretical models
because it is still a big challenge for some ab initio models
including lattice QCD to reproduce the saturation properties,
starting from fundamental particle interactions. 
This is also demonstrated in \S\ref{sec:examples}.
The consistency of the saturation energy for phenomenological models,
on the other hand, should be ensured because these models
are determined to reproduce those properties.

\subsection{Symmetry Energy}\label{sec:sym}
The symmetry energy \esym and its slope $L$ are expressed as a parameter $a_4$ that is
a representative around the saturation density. The incompressibility $K$ is also given in the similar expansion
of the energy of symmetric nuclear matter.
The EOSDB treats these properties as a function of 
baryon density, which enables to check the compiled EoSs explicitly.
These constraints (\esym, $L$, and $K$) 
are useful in the comparisons of the data taken from different papers.

The symmetry energy is defined in terms of a Taylor series expansion of the energy for nuclear matter as a function of the 
charge asymmetry $\delta = 1 - 2 Y_{\rm C}$,
\begin{eqnarray}
E(\rhob, \delta) &=& E(\rhob, 0) + \frac{1}{2!}\frac{\partial^2 E(\rhob,\delta)}{\partial\delta^2}|_{\delta=0}\delta^2
+ \frac{1}{4!}\frac{\partial^4 E(\rhob,\delta)}{\partial\delta^4}|_{\delta=0}\delta^4 + {\cal O}(\delta^6) \\
a_4 &=& \frac{1}{2!}\frac{\partial^2 E(\rhob,\delta)}{\partial\delta^2}|_{\delta=0}
= a_4(\rho_0) + L\epsilon + {\cal O}(\epsilon^2) \\
E(\rhob,0) &=& E(\rho_0,0) + \frac{K}{2!}\epsilon^2 + {\cal O}(\epsilon^3) \\
K &=& 9\rho_0^2\frac{\partial^2 E_o(\rhob)}{\partial\rho^2}|\rhob=\rho_0\\
\epsilon &=& (\rhob -\rho_0)/3\rho_0
\end{eqnarray}
where $\rho_{B}$ and $\rho_0$ denote the baryon density and the saturation density of nuclear matter, respectively.
If the $a_4$ is not given in the literature,
we compile the approximated form of the symmetry energy,
i.e., energy difference between
symmetric nuclear matter and pure neutron matter at arbitrary baryon densities.
 
The symmetry energies are available at any given baryon densities from theoretical models
in the lieterature, while the availability is limited for experimental data.
The compiled data are also to be compared with experimental data.
In the current experimental setups, the symmetry energy is measured for the limited
range of the saturation density, ($\sim 1/3 - 1$ times $\rho_{0}$).
However, ongoing and future experiments
are expected to determine the symmetry energy at higher densities up to $\sim 3 \rho_{0}$.

\section{Search and Plot System of the EOSDB}\label{sec:system}
We have constructed a database subsystem for use of the EOSDB.
Users can access and select data based on various criteria, some of which are
shown in Table~\ref{tab:quantity}.
The selected data can be drawn in the viewgraph on the browser with user-specified axes,
and the results can be downloaded from the server.
All the data are linked to the text files and original papers
where required information are available.
 
\subsection{Data Search}
\begin{figure}
  \begin{center}
    \FigureFile(160mm,160mm){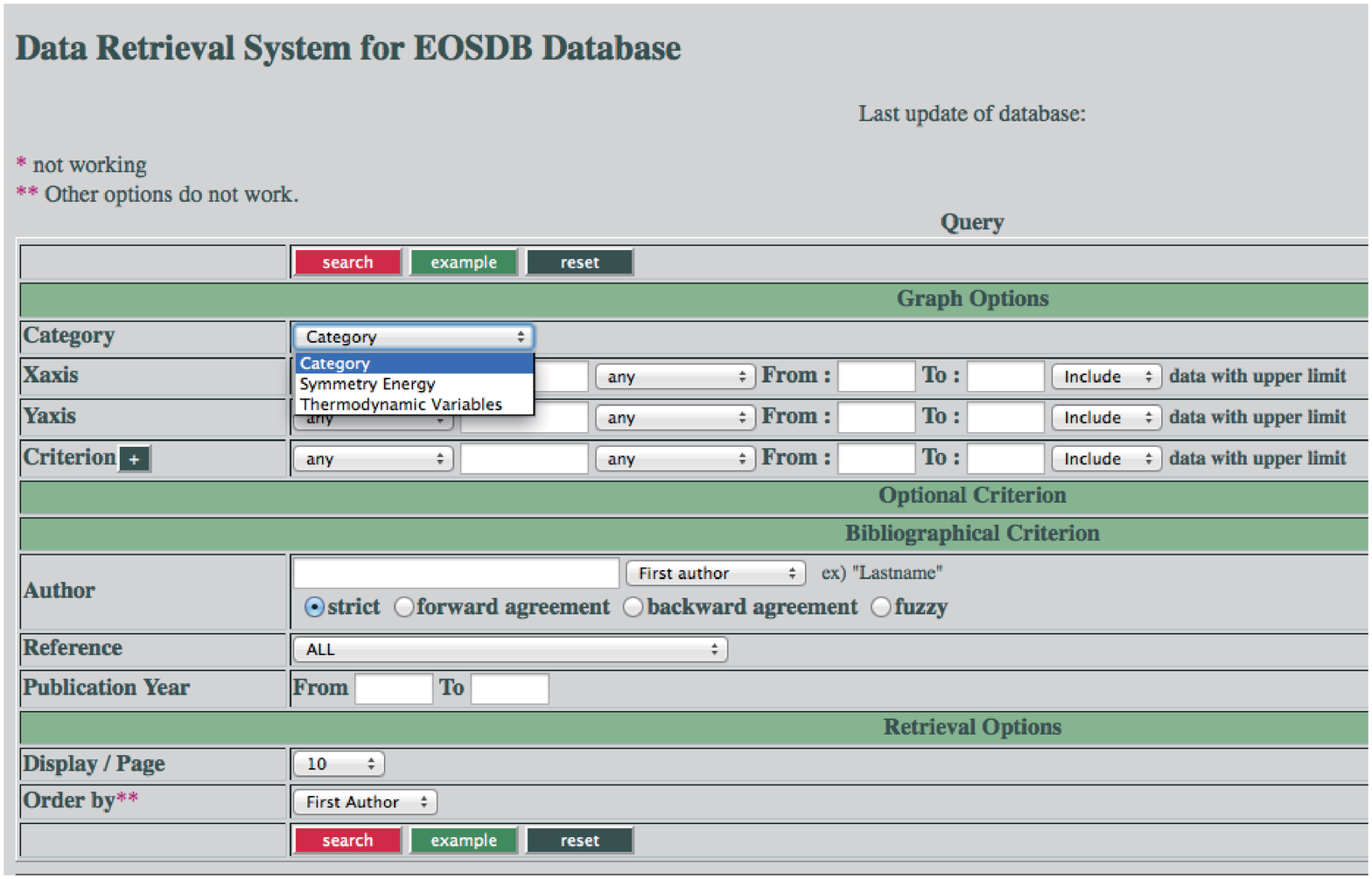}
  \end{center}
	\caption{
Screen snapshot of the top page of the search and plot system for the EOSDB. 
	}
	\label{fig:EOSDB_step1}
\end{figure}

Figure~\ref{fig:EOSDB_step1} shows the screen snapshot of the search and plot system.
Search criteria are divided into three sections in the form of the system.
The first section of the form provides criteria related to the axes of the diagram.
The first line specfies the category of the search.
In the current version, users can specify ``Symmetry Energy'' or ``Thermodynamic Variables'',
which is used to help to specify the axes of the graph depending on the properties of EoSs,
but this option is under construction and will be used in the future update.
The next two lines are used to specify the quantities to draw in the graph. 
Users can select from the following quantities for each axis in the first column:
baryon density, $\rhob$, symmetry energy, $\esym$, slope coefficient, $L$,
incompressibility, $K$, energy, pressure, and entropy.
Users can also enter one of these quantities in the text box in the second column:
All these variables are given as a function of $\rhob$ in the database.
Therefore, the default option for the quantity in the X-axis is set at $\rhob$.
Those who are interested in symmetry energy should choose $\esym$, $K$, or $L$ for the Y-axis, 
while users interested in thermodynamic variables should choose energy, pressure, or entropy.
The third column gives an option to specify the value of the charge ratio ($\yc = 0.0$ or $0.5$).
This should be specified unless the graph axis is set at $\rhob$.
As described in \S~\ref{sec:sym}, the option, $\yc = 0.0$ means pure neutron matter
or charge neutral matter.
It is recommended to check which condition is realized in compiled data by tracing the
links to individual text data as described in \S~\ref{sec:chardb}.
The 4th and 5th columns set the range of the values for each selected quantity, with the
option in the 6th column of whether to include or exclude data that report the quantity 
with only an upper limit.
In the 4th line, users can specify the required range in the data, if necessary,
to select or remove the data from plotting, e.g., by setting, $0 \leq E/B \leq 500$~MeV.
The number of criteria can be extended to as many as desired by the user.

The second section of the form is used to extract specified papers.
Through the use of these options, one can extract the data containing
a specific author, journal, and the range of the year of publication.

Retrieval options are set in the third section, such as the number of data
to display in the resulting list and order of the list.

\begin{figure}
  \begin{center}
    \FigureFile(80mm,80mm){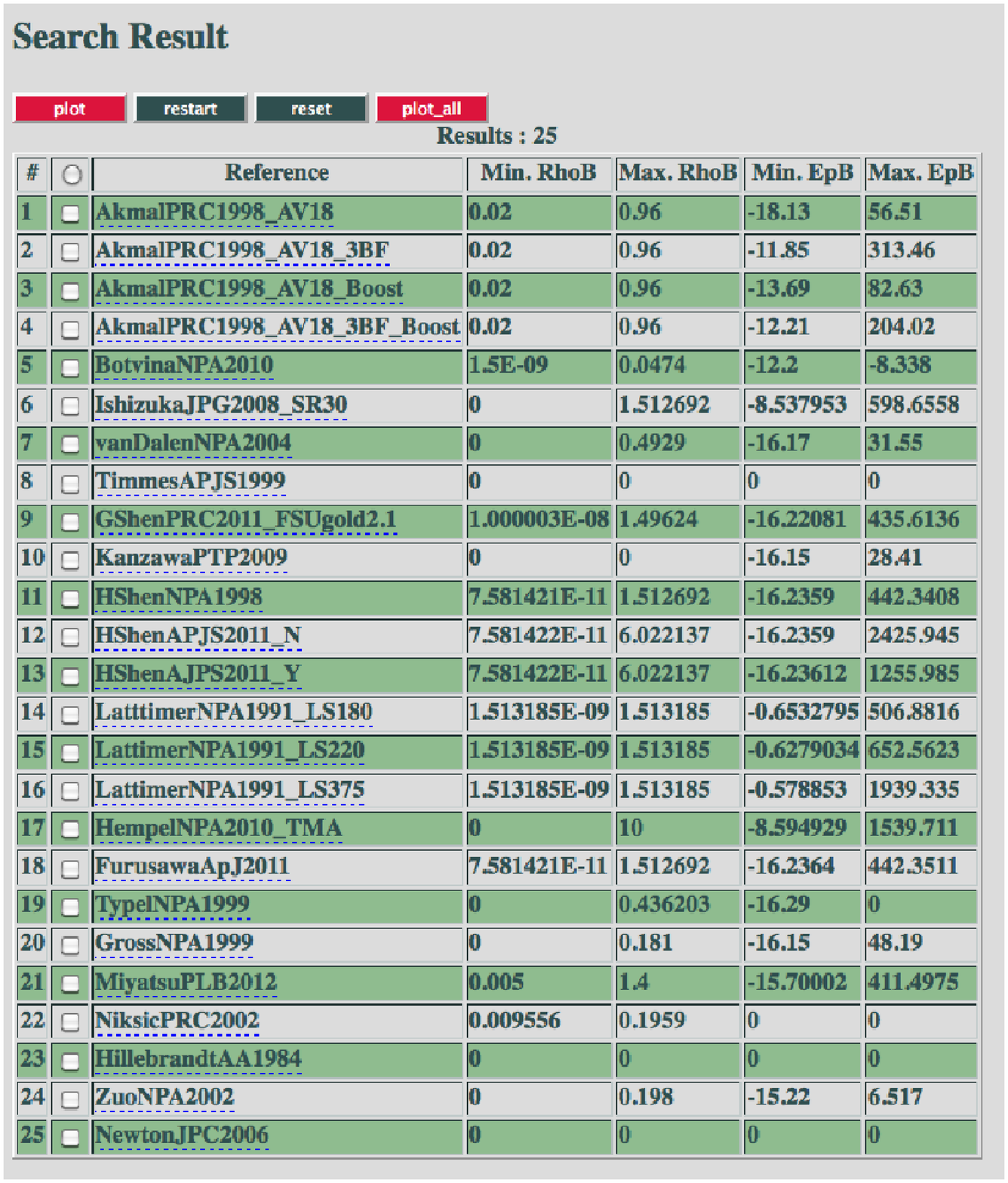}
  \end{center}
	\caption{
Screen snapshot of the search result of the search and plot system of the EOSDB.
In this case, the X-axis and Y-axis are set to $\rhob$ and $E/B$ (energy per baryon), respectively,
for the category of thermodynamic variables.
See text for the meanings of the columns in the table.
	}
	\label{fig:EOSDB_step4}
\end{figure}

\begin{figure}
  \begin{center}
    \FigureFile(160mm,160mm){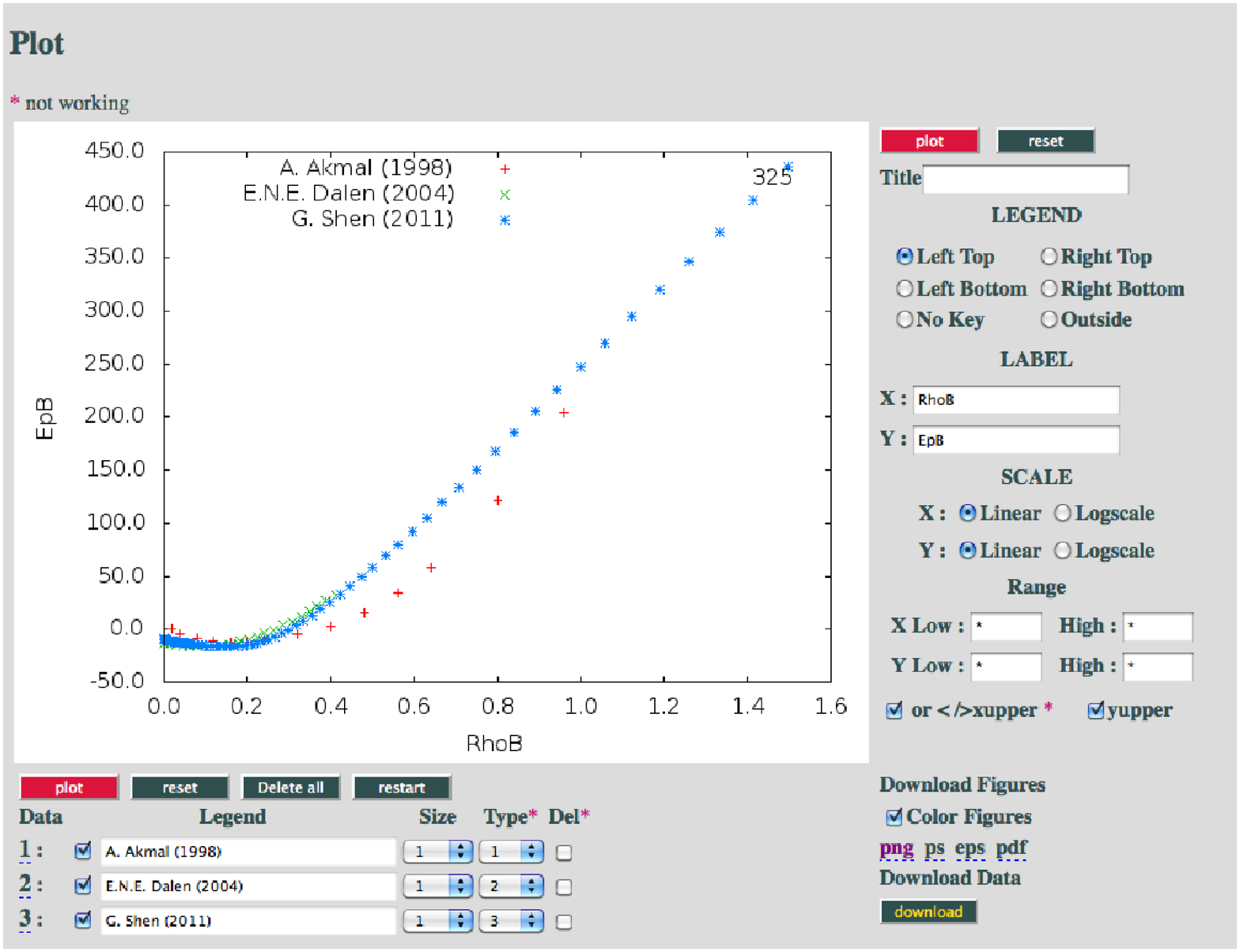}
  \end{center}
	\caption{
Screen snapshot of the data plotting of the search and plot system of the EOSDB.
Three sets of data were selected from the list shown in Fig.~\ref{fig:EOSDB_step4}.
See text for the equipments in the system and the options to edit the diagram on the browser.
	}
	\label{fig:EOSDB_step5}
\end{figure}

A screen snapshot of an example for the retrieved set of records
is shown in Figure~\ref{fig:EOSDB_step4}.
The retrieved records are displayed in table format on the browser.
The columns represent, from left to right, the checkbox to select data
to be plotted, the reference ID, and the minimum and maximum values for the
quantities selected as the X-axis and Y-axis of the plotted diagram, respectively.
By using the provided links to the reference ID, one can trace the information
on the data stored as text format as listed in \S~\ref{sec:chardb}.
For selected data, the diagram is drawn in the web browser according to
the choice of options, using the publicly available graphic software Gnuplot
(see Figure~\ref{fig:EOSDB_step5}).
Graphs drawn in the browser are equipped with simple functions for editing.
The standard options are to change the labels, position of the legend, and the scales
and ranges of the graph.
Users can also download the figures in various formats (png, eps, ps, and pdf,
in color or in black and white).
Plotted numerical data, as well as the script to reproduce the figure on
the screen can be downloaded from the server, if one wishes to edit the graph
in more detail.
Numerical data are also accessible by tracing the link to each data set
in the list.
Users can upload their own data to the server to compare them with the
plotted data.

It is recommended to refer the detailed information when one compares
the plotted data with the system. In particular, assumptions and methods
adopted in theoretical models should be checked so that the comparison
is based on appropriate conditions. In the future update of the system,
this information will be added as a criterion for choosing the data.

\section{Application to Model Analysis} \label{sec:examples}
We demonstrate how the data in the EOSDB can be used to analyze theoretical models.
First, the EoSs widely used in astrophysics community are examined
using the symmetry energy registered in the EOSDB.
Second, theoretical models are compared using the density dependence of energy.

\begin{figure}
  \begin{center}
    \FigureFile(80mm,){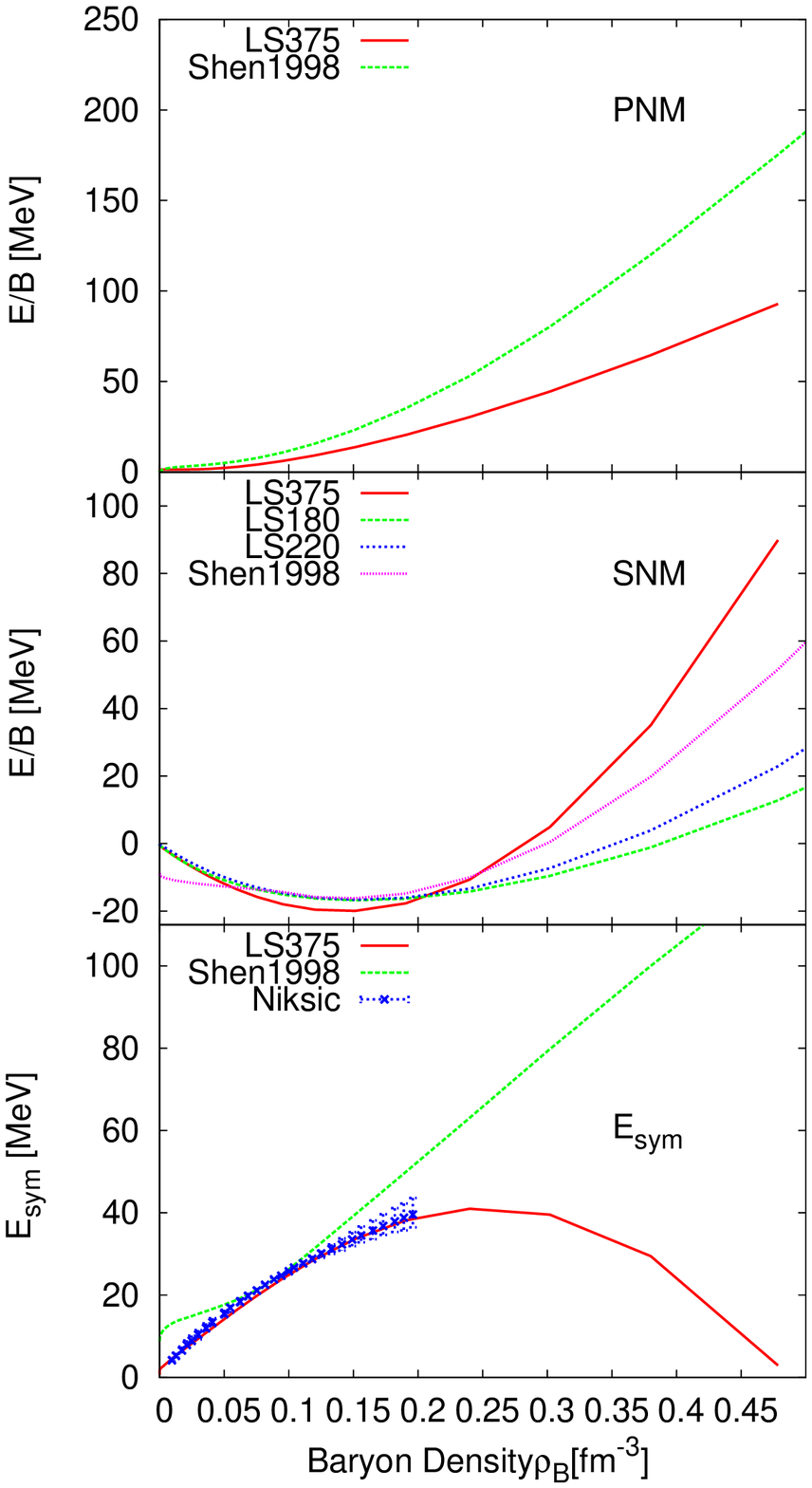}
  \end{center}
 	 \caption{
Comparisons of energy per baryon or symmetric energy as a function of baryon density for different sets of nuclear EoSs.	 
The upper and middle panel show the energy density of pure neutron matter and symmetric nuclear matter, respectively.
Note that Shen EoS includes inhomogeneous phase at lower densities than the saturation, while the others are
calculations for simple uniform matter. 
In the upper and lower panel, the red solid line is Lattimer-Swesty EoS with the incompressibility of $K = 375$ MeV, while the green dashed line represents Shen EoS only with nucleons. In the middle panel, the three options of Lattimer-Swesty EoS are plotted in the red solid line ($K = 375$ MeV), the greed dashed line ($K = 220$ MeV) and the blue dotted line ($K = 180$ MeV) in comparison with Shen EoS. The blue error bar in the lower panel shows an experimental error bars. 
	}
	\label{fig:TM1-LS220}
\end{figure}

In Figure \ref{fig:TM1-LS220}, we compare energy per baryon ($E/B$) and symmetry energy
as a function of baryon densities using two EoSs, Lattimer \& Swesty's EoS \citep[hereafter LS EoS]{LS} and Shen's EoS \citep[hereafter Shen EoS]{Shen1998}, 
both of which are commonly used in astrophysical studies of neutron stars, supernovae and black holes. 
These data are also compared with experimental data in the bottom panel.
Here, the dataset {\it LS180}, {\it LS220}, and {\it LS375} denotes the LS
EoS at the incompressibility of $180$, $220$, and $375$ MeV, respectively.

Each dataset is obtained by the following procedure.
The LS EoS has been compiled using the analytic equations in \citet{LS} that describe energy as a function of 
baryon densities at $T = 0$ MeV.
It is to be noted that this EoS is only for uniform matter. 
This is due to a problem in computing an EoS using the distributed version of the program
that was supposed to give a table with nucleons including leptons and photons at low temperatures and low $\yc$.
The {\it LS375} provides the best consistency with the parameter set for Skyrme force,
whose symmetry energy is consistent with experimental constraints.
The Shen EoS data is taken from their EoS tables using the RMF parameter
set, {\it TM1}. In the tables, the inhomogeneous phase under $\rho_0$ 
is treated by Thomas-Fermi approximation.
The EoSs with this parameter set gives generally lower energy than those for uniform matter.
In both types of EoSs, the symmetry energy is defined as the energy difference
between pure neutron matter and symmetric nuclear matter.
We compare the dataset ``Niksic (2002)'' with LS and Shen EoSs in the bottom panel of Fig.~\ref{fig:TM1-LS220}.
The data provides the constraints on EoS from experiments and are compiled from \citet{Fuchs2006} using the GSYS. 
We represent the constraint on the symmetry energy with error bars, while it is shown as a shaded area in the original figure.
The constraint on the symmetry energy is obtained experimentally from $^{208}$Pb and $\alpha$ inelastic 
scattering data for Iso-Vector Giant Dipole Resonance using a density dependent relativistic mean field (DDRMF) parameter set, 
{\it DD-ME1} and {\it PRA} for exited modes.
As shown in Table~\ref{tab:compiled_data},
the basic properties of these EoSs are (E$_{\rm sym}$, $K$, \nsmax) 
$=$ ($29.3$ MeV, $220$ MeV, $2.06 M_{\odot}$) for {\it LS220} and ($37.9$ MeV, $281$ MeV, $2.18 M_{\odot}$) for {\it TM1}.
These are in good agreement with the recent experimental constraints on the value of $K$ of $230 \hyp 270$ MeV.

The large discrepancy between these two EoSs can be understood as follows, speculated from the differences in their theoretical models.
The major difference in models between LS EoS and Shen EoS is the condition assumed for a nuclear system.
In the LS EoS, a modified Skyrme I force (SkI) \citep{Vautherin1970} is used. 
The SkI can reproduce the properties of closed shell nuclei such as $^{16}$O, $^{40,48}$Ca, $^{90}$Zr, $^{208}$Pb.
They adjusted the incompressibility by adding a three-body interaction parameter.
Adding the three-body interactions in the Skyrme forces can be justified only if they can reproduce experimental values. 
The dataset {\it TM1}, on the other hand, is produced by a parameter set of relativistic mean field (RMF) that is adjusted to reproduce
both the binding energies and the charge radii of proton-rich and neutron-rich nuclei as well as representative closed shell nuclei
such as $^{8 \hyp 20}$C, $^{14 \hyp 22}$O, $^{28, 34}$Si,
$^{40, 48}$Ca, $^{90}$Zr, $^{116,124}$Sn, and $^{184 \hyp 214}$Pb.    
The RMF models
naturally involve the relativity effect that is known to make an EoS stiffer
than non-relativistic EoSs.
Shen EoS covers nuclear matter at high densities and various $\yc$,
which is useful in applications to astronomical phenomena such as supernovae and the formation of neutron stars.

It is also to be noted that there is a limitation in the application of the Skyrme Hartree Fock and RMF models.
Both the Skyrme Hartree Fock and the RMF models are
based on experimental analyses of the HIC data to constrain the symmetry energy and its slope with respect to $\rhob$.
The Skyrme Hartree Fock models can well describe various finite nuclei at low energy,
although it should be applied below $E/B < 50$ MeV
because it is difficult to determine a Skyrme parameter that can reproduce both Pb and Sn at the same time \citep{Stone2003}.
This appears to be in conflict with the fact that heavy ion collisions at high energies are required
to derive the symmetry energy above $\rho_0$.
On the other hand, the RMF models can explain p-induced reactions even at high energies but
it shows the poor reproduciblity of experimental data such as binding energy and charge radius for light nuclei.
We should also note that the RMF includes only direct interaction,
and that the exchange interaction (the Fock term) might be necessary in a dense many-body system. 
Thus these major models have their advantages and disadvantages.
However, Skyrme Hartree Fock models have been widely used in the analysis for the symmetry
energy thanks to their plentiful variety.
Some Skyrme forces have a peak in the symmetry energy at around the saturation density,
the others show almost mono-topical increase of $E_{\rm sym}$ as the density rises,
while that of RMF models mono-topically arises as a function of the density, in general.
As shown in the bottom panel of Fig.~\ref{fig:TM1-LS220}, there is an increasing discrepancy of $\esym$ 
with increasing density between these two models. 
Experiments to constrain the symmetry energy are ongoing in such a high $\rhob$ region. 

The above discussion tells us why we need careful treatments
on the saturation property and symmetry energy. 
At around $\rhob = 0.1$ fm$^{-3}$, 
both LS and Shen EoSs show a reasonable agreement with the experimental constraint
on the symmetry energy.
Especially at around $\rho_0$, {\it LS375} satisfies the constraint. 
However, it has too large value of the incompressibility compared with
that constrained by experiments, i.e., 230 $\hyp$ 270 MeV.
As for the other datasets, {\it LS220} and {\it LS180},
they also show reasonable agreement with the symmetry energies around the saturation density. 
However, they do not provide the best fit with the result with the SkI' force,
which is fitted to reproduce various closed shell nuclei with its incompressibility $K = 370$ MeV.
In addition, the incompressibility is smaller for {\it LS220} and {\it LS180} than that constrained by experiments.
The saturation property of Shen EoS is similar to the softer EoSs like {\it LS220} and {\it LS180}.
In Shen EoS, the symmetry energy seems to be larger in a few MeV than 
the constraint around $\rho_0$,
even though it agrees well with the recent experimental constraint
for the symmetry energy, 31 $\pm$ 3 MeV at the saturation density, $\rho_0$.
Its incompressibility $K = 281$~MeV is also slightly large compared with 230 $\hyp$ 270 MeV.
In conclusion, it is difficult to satisfy the constraints on both the incompressibility and
the symmetry energy simultaneously for the EoSs widely accepted by astrophysics community.
It may be caused by the model dependence contained in the constraint itself,
because the experimental analysis has been performed with various models
for which the analysis should not be applied.

\begin{figure}
  \begin{center}
    \FigureFile(160mm,160mm){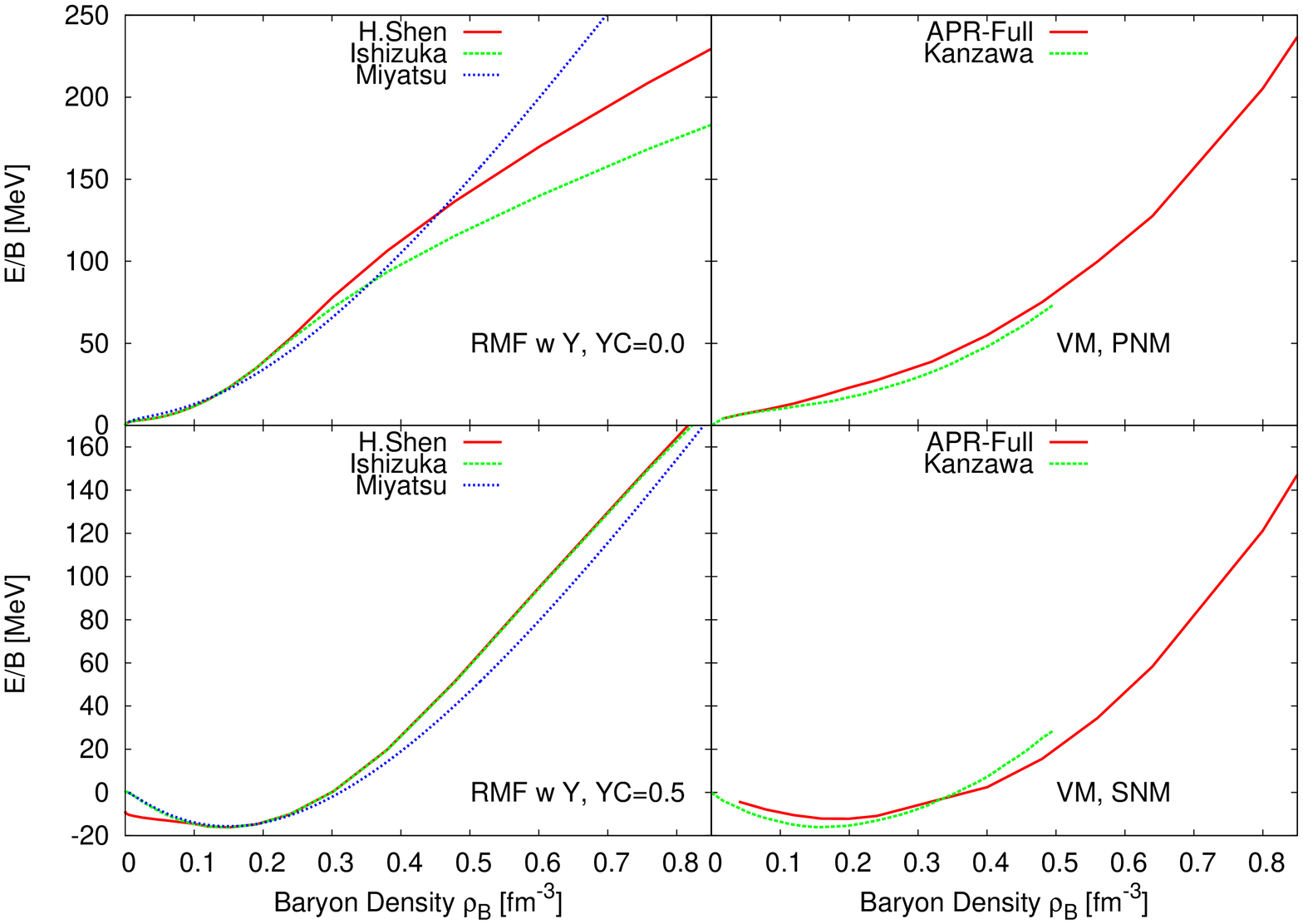}
  \end{center}
 	 \caption{
Comparison of ab initio EoSs and phenomenological relativistic EoSs using the data downloaded from the EOSDB data retrieval system.
	}
	\label{fig:RMF-VM}
\end{figure}

Figure~\ref{fig:RMF-VM} demonstrates another advantage of using the EOSDB.
It compares energy as a function of baryon densities for results with different theoretical models.
We present the models of the RMF that is based on a phenomenologial framework,
and those of variational methods that is based on ab initio calculations.

Along with the different characteristics of theoretical models, the compiled models of nuclear matter can
be divided into two groups as shown in Table~\ref{tab:Classification_1}, \ref{tab:Classification_2} and \ref{tab:Classification_3}.
Each table has a list of models and characteristics, and the mass and radius of neutron stars
together with the corresponding data ID and the reference ID in our database.
In both tables, the symbol \nsmax  in the last column is the maximum mass of neutron stars in each model,
and $R$ is the radius at the \nsmax.
We should note that the radius could vary according to treatment of the neutron star crust.
We calculated the radii of E0002 and E0012 using the Shen EoS table for the crust.
As for the other entries,
the detailed information on the crust treatment can be seen in the references ~\citep{NSR-1,NSR-2,NSR-3,NSR-4}.

In Table~\ref{tab:Classification_1},
the first column denotes the adopted framework which is either relativistic or non-relativistic.
The acronyms ``VM'', ``BHF'', and ``DBHF'' in the second column mean
Variational Method, Brueckner-Hartree-Fock, and Dirac-Brueckner-Hartree-Fock, respectively. 
The label ``$NN$'' in the third column denotes nucleon-nucleon interaction while
``$NNN$'' in the last column means the three-body interaction. 
In Table~\ref{tab:Classification_2} and \ref{tab:Classification_3},
the first column denotes the adopted framework which is either relativistic or non-relativistic.
The acronyms
``RDBHF'', ``SKF'', ``RMF'', ``RHF+QMC'' and ``DDRMF'' represent the method of calculations and correspond to Skyrme Hartree-Fock,
Relativistic Mean Field, Relativistic Hartree Fock with Quark Meson Coupling Model and Density Dependent RMF, respectively. 

In the left panels of Figure~\ref{fig:RMF-VM}, we present
three models with the RMF theory with hyperons ($Y$) taken into account and have compiled in the EOSDB.
These three models have the following characteristics.
The Data labeled ``H.Shen'' \citep{Shen2011} contains only $\Lambda$ as hyperons 
using the RMF parameter set {\it TM1}.
In the data labeled with ``Ishizuka'' \citep{IOS2008}, we use repulsive $30$ MeV case of $\Sigma N$ interaction model,
which gives a good agreement with $^{28}$Si($\pi^-$, $K^+$)-reaction of the KEK E438 experiment
\citep{Maekawa2007}.
We calculated a new dataset to remove the contribution by leptons, but
omitted the inhomogeneous phase for simplicity.
These two EoSs are based on the same parameter set {\it TM1} for nuclear part.
As shown in the bottom left panel, they show the same behavior with each other
in the $Y_C$ = 0.5 case.
On the contrary,
the difference between these models increase with the density in the $Y_C$ = 0 case (top right panel).
This is because the Ishizuka EoS contains more neutral $n$ and $\Lambda$ than Shen EoS
due to the inclusion of the other charged hyperons.

We display another EoS with hyperons to examine the possibility to distinguish the different constituents
within multi-theoretical frameworks when an EoS, which is determined from observational
masses and radii of neutron stars, is provided.
As seen from the top left panel of Fig.~\ref{fig:RMF-VM}, the discrepancy caused by constituents is
more significant than that caused by different parameters at $\rhob \gtrsim 3 \rho_{0}$.
The data of ``Miyatsu'' \citep[private communication]{Miyatsu2012},
are based on the calculations using a relativistic Hartree-Fock method with a quark-meson coupling model.
It also contains full baryon octet as well as the Ishizuka EoS.
The data of Miyatsu and Ishizuka show similar behavior below the saturation density as shown in the left panels,
while they do not agree with the result of Shen EoS.
This is because the inhomogeneous phase in Shen EoS gives lower energy in a system than the others that are based on uniform matter calculations. In fact, Shen EoS and Ishizuka EoS are consistent with each other 
at higher densities than the saturation in the case of nucleon matter.
Another difference between Miyatsu and the other RMF EoSs is the Fock term in high densities,
which is neglected in the RMF models.
This effect gives a stiff EoS enough to support a massive neutron star, which is also consistent with the Shapiro delay observations \citep{Demorest2010}  as seen in the left upper panel.

In the right panels of Fig.~\ref{fig:RMF-VM}, we compare two ab initio calculations.
The data labeled with ``APR-Full'' is taken from tables shown in \citet{APR1998}.
The APR EoS is the representative of ab initio calculations based on the fundamental interactions of
nuclear many body systems.
They use the variational method with two- and three-body interactions and Lorentz boost for relativistic correction.
The data labeled with ``Kanzawa'' \citep{Kanzawa} follows the APR EoS scheme whose data are scanned from the viewgraphs.
In both cases of pure neutron matter (the top right panel) and
symmetric nuclear matter (the bottom right panel), 
these models show almost the same properties with each other,
except for the small gap.
As for the APR data, we adopt the values before adjusting its binding energy to the empirical value of $-16$ MeV
at $\rhob=0.16 $fm$^{-3}$ in order to include the many-body corrections and the other corrections to their EoSs, separately.
These information are necessary to estimate the influence of each component on EoSs.
On the other hand, the Kanzawa EoS is the data with the adjustment
because data only with the correction are provided in the paper.
The correction term in the Kanzawa EoS satisfies the same condition required in the APR EoS. 
From the comparison of these EoSs with and without the correction of the binding energy,
we need to pay enough attention to a criterion of each data in collecting figures from tables in published articles.
At present, such saturation property calculated by ab-initio models has been used as a fitting condition
even in the other ab initio calculations such as lattice QCD.
To derive the saturation property from experiments has a mutual dependence with nuclear models, similar to the case for symmetry energy.
Such dependences are inevitable for ab-initio calculations.
This problem is expected to be resolved by 
the improvement of experimental techniques and
high performance computers.

In summary, we demonstrated how the difference among theoretical models,
assumptions and constituents affect the basic properties of nuclear matter
using visualization tools of the EOSDB.
In addition, exploring the relationships between two physical quantities may
lead us to find important diagnoses to constrain EoSs in different viewpoints. 
Moreover, the visualizations reveal the overall picture of the model dependence of various EoSs.
In Fig.~\ref{fig:RMF-VM},
it has been confirmed that the RMF models give stiffer EoSs than the non-relativistic 
ab initio EoSs, especially above the saturation density $\rhob \sim 0.16$ fm$^{-3}$.
It has also been confirmed that
the incompressibility in the RMF models is larger than that in the ab initio models.
From the comparison between the left and right panels,
we find that the present models agree well with these patterns
by checking energy dependence on the density in the symmetric nuclear matter and 
the curvature at $\rho_0$.

\begin{landscape} 
 \begin{table}
  \caption{Table for classification of ab initio theoretical models. 
  }\label{tab:Classification_1}
  \begin{center}
    \begin{tabular}{llllll}
     \hline    
     Ab initio & & & & & \\
      \hline       
      Rel. / Non-rel. & Method & Interaction & Reference                & Data ID & Comment \\         
      \hline       
      Non-rel. & VM    & AV18 for NN   & AkmalPRC1998                   & E0016 & (\nsmax, R) = (1.67M$_\odot$, 8.2 [km]). \\  
      Non-rel. & VM    & AV18 for NN   & AkmalPRC1998\_AV18\_3BF        & E0017 & UIX for NNN. (\nsmax, R) = (2.38M$_\odot$, 10.08 [km]).  \\    
      Non-rel. & VM    & AV18 for NN   & AkmalPRC1998\_AV18\_Boost      & E0018 & Relativistic Correction included \\
               &       &               &                                &       &  (\nsmax, R) = (1.80M$_\odot$, 8.75 [km])\\  
      Non-rel. & VM    & AV18 for NN   & AkmalPRC1998\_AV18\_3BF\_Boost & E0019 & UIX for $NNN$.  \\
               &       &               &                                &       & Relativistic Correction included \\   
               &       &               &                                &       &  (\nsmax, R) = (2.20M$_\odot$, 10.04 [km])\\  
      Non-rel. & VM    & AV18 for NN   & KanzawaPTP2009                 & E0010 & UIX for NNN. \nsmax = 2.2M$_\odot$.  \\
      Non-rel. & BHF   & AV18 for NN   & ZuoNPA2002                     & E0020 & Tuscon-Melbourne for $NNN$  \\
      Rel.     & DBHF & Bonn-A for NN & Gross1999                      & E0021 & Symmetric Matter \\
               &       &               &                                &       &  (\nsmax, R) = (2.24M$_\odot$, 10.78 [km]) for ($n,p,e^-$) matter\\       
      Rel.     & DBHF & Bonn-A for NN & vanDalenNPA2004                & E0022 & Asymmetric Matter \\     
                &       &               &                                &       &  (\nsmax, R) = (2.24M$_\odot$, 10.78 [km]) for ($n,p,e^-$) matter\\                         
      \hline
   \multicolumn{6}{@{}l@{}}{\hbox to 0pt{\parbox{160mm}{\footnotesize
     }\hss}}
    \end{tabular}
  \end{center}
\end{table}
\end{landscape} 

 \begin{landscape} 
 \begin{table}
  \caption{Table for classification of phenomenological theoretical models.
  }\label{tab:Classification_2}
  \begin{center}
    \begin{tabular}{llllll}
     \hline    
    Phenomenological  & & & & & \\          
      \hline       
 Rel. / Non-rel. & Method      & Interaction & Reference                & Data ID & Comment \\         
      \hline       
       Non-rel. & SHF          & SkI'        & LatttimerNPA1991\_LS180  & E0005 & $K$=180.  \\   
                 &             &               &                        &       &  (\nsmax, R) = (1.84M$_\odot$, 10.2 [km]). \\      
       Non-rel. & SHF          & SkI'        & LatttimerNPA1991\_LS220  & E0006 & $K$=220. \\
                  &             &               &                        &       &  (\nsmax, R) = (2.06M$_\odot$, 10.85 [km]). \\ 
       Non-rel. & SHF          & SkI'        & LatttimerNPA1991\_LS375  & E0007 & $K$=375. \\
                  &             &               &                        &       &   (\nsmax, R) = (2.72M$_\odot$, 12.53 [km]). \\   
       Non-rel. & 3Dim. SHF    & SKa         & HillebrandtAA1984        & E0013 & Data shown only in Entry.html \\
                &             &               &                        &       & (\nsmax, R) = (2.21, 11.7 [km]).  \\
       Non-rel. & 3Dim. SHF    & SkM$^*$     & NewtonJPC2006            & E0015 & --- \\   
       \hline      
          \multicolumn{6}{@{}l@{}}{\hbox to 0pt{\parbox{160mm}{\footnotesize
     }\hss}}
    \end{tabular}
  \end{center}
\end{table}
\end{landscape} 

 \begin{landscape} 
 \begin{table}
  \caption{Table for classification of phenomenological theoretical models.
  }\label{tab:Classification_3}
  \begin{center}
    \begin{tabular}{llllll}
     \hline    
    Phenomenological  & & & & & \\          
      \hline       
 Rel. / Non-rel. & Method      & Interaction & Reference                & Data ID & Comment \\         
      \hline       
       Rel.     & RMF          & TM1(Only N) & HShenNPA1998             & E0002 & Thomas-Fermi apprx.  \\      
                &              &             &                          &       &  for inhomo. phase. \\
                &             &               &                        &       & (\nsmax, R) = (2.18M$_\odot$, 12.5 [km]).\\
       Rel.     & RMF          & TM1(Only N) &  HShenAPJS2011\_ N       & E0003 & Different from E0002 at $(T, Y_p) = (0, 0)$. \\
                &              &             &                          &       &  (\nsmax, R) = (2.18M$_\odot$, 12.5 [km]). \\
       Rel.     & RMF          & TM1(Only N) & FurusawaApJ2011          & E0011 & NSE for inhomo. phase \\
       Rel.     & RMF          & TM1(Only N) & BotvinaNPA2010           & E0010 & NSE for inhomo. phase \\
       Rel.     & RMF          & TM1(with Y) & HShenAPJS2011\_Y         & E0004 & Only $\Lambda$ included as hyperons. \\
                &              &             &                          &       & \nsmax = 1.75M$_\odot$. \\
       Rel.     & RMF          & TM1(with Y) & IshizukaJPG2008\_SR30    & E0012 & Full Baryon Octet. \\
                &             &               &                        &       & (\nsmax, R) = 1.63M$_\odot$, 13.26 [km]). \\
       Rel.     & RMF          & TMA         & HempelNPA2010\_TMA       & E0008 & NSE for infomo. phase \\
                 &             &               &                        &       & (\nsmax, R) = (2.04M$_\odot$, 12.43 [km])\\
       Rel.     & RMF(RHF+QMC) & ---         & MiyatsuPLB2012           & E0009 & Full Baryon Octet. \nsmax = 1.95M$_\odot$. \\
       Rel.     & DD RMF       & DD-TW       & TypelNPA1999             & E0023 & (\nsmax, R) = (2.2M$_\odot$, 11.2 [km]). \\
       Rel.     & DD RMF       & DD-ME1      & NiksicPRC2002            & E0024 & (\nsmax = 2.47M$_\odot$, 11.9 [km]). \\
       Rel.     & DD RMF       & FSUgold     & GShenPRC2011\_FSUgold2.1 & E0001 & Adjusted to support 2.1M$_\odot$ NS.\\ 
                &              & + Polytrope &                          &       & (\nsmax, R) = (2.1M$_\odot$, 12.2 [km]) \\
       \hline      
          \multicolumn{6}{@{}l@{}}{\hbox to 0pt{\parbox{160mm}{\footnotesize
     }\hss}}
    \end{tabular}
  \end{center}
\end{table}
\end{landscape} 

\section{Summary and future prospects}

We have constructed the database of nuclear EoSs (EOSDB), which is available online.
The database includes information on experimental or theoretical details,
energy, pressure, entropy, symmetry energy, the derivative of the symmetry energy 
with respect to baryon density, and incompressibility as functions of baryon density.
These data are taken from published papers, with the help of a software to
scan viewgraphs.
A search and plot system has been
converted from the SAGA database that deals with observed metal-poor stars in the Galaxy.
The system enables the retrieval and plotting of data selected according to
various criteria featuring nuclear saturation properties.
Our sample includes 36 datasets mainly
for symmetric nuclear matter, pure neutron matter and its constraints (the symmetry energy) at $T = 0$ MeV.
The list of the compiled data are presented in tabular format.
The summary of the theoretical models compiled in our database together with the derived maximum mass of neutron stars is also presented
according to the classification of theoretical models.

The EOSDB can help to examine various EoSs because the data is provided in a unified format.
However, the users should note that these data are based on different assumptions and models
and may cause problems in attempting comparisons among EoSs without understanding their details.
One of the future updates will include the query options according to models and methods in the search and plot system.
This will elucidate the model dependence of EoSs and the physics behind
qualitatively and quantitatively
in more efficient way.
We will report in a forthcoming paper a benchmark
test for various EoSs using the EOSDB.
Another future update will be to try to receive data from the authors of the papers,
instead of scanning viewgraphs, to guarantee the quality of the compiled data.

We demonstrated the model dependence of EoSs using the EOSDB and find that
theoretical EoSs commonly used in astrophysics have a difficulty in satisfying
the experimental constraints on both the incompressibility
and the symmetry energy simultaneously.
This suggests that we need more sophistications of models that deal with
nuclei.
For example, a compound system of baryons, which spans a wide range
of size and energy, should be treated as static or dynamical context.

The EOSDB can be an even more powerful tool with the help of the future observations of neutron stars.
In this paper, we also summarize the theoretical mass and radius of a cold neutron star, although
it could depend on the treatment of neutron star crust. We will soon report the details of the dependence.
It is currently very difficult to measure the mass and radius of neutron stars simultaneously;
For isolated neutron stars or magnetars, the surface temperature and radius (and possibly magnetic field) 
can be measured, while their masses cannot be determined in such systems.
The mass can be determined with uncertainties associated with the inclination angle of the orbital plane of binary neutron stars.
A possible case
to measure the mass and radius will be for X-ray binaries with weak magnetic field.
In this case, the information of the distance (or redshift), the temperature, and possibly the surface gravity of neutron stars are required.
Still, we should keep in mind that these values contain ambiguities in absorption lines used, the assumption of
blackbody radiation, and atmosphere models.
Therefore, to increase the quality of the observed parameters, we need more sample to compare,
which will be achieved by future observations of neutron stars.
The ASTRO-H, which is scheduled for the launch in 2015, will enable us to
analyze the 4.1keV absorption line of the neutron star X-ray transient 4U 1608-52 thanks to its high resolution spectra.
The observations of this object was so far performed only with Tenma. 
The Neutron star Interior Composition ExploreR (NICER), which will be launched in 2016,
enables rotation-resolved spectroscopy of the thermal and non-thermal emissions of neutron stars.
In the early 2020s, the Large Observatory for X-ray Timing (LOFT) is also proposed.
The EOSDB has a potential to include these data in the future updates.
We are planning to compile more detailed information about neutron stars such as binarity, mass, radius, and magnetic field for thousands of neutron stars that have already been observed in our Galaxy, 
which may give us an opportunity to discuss the EoS and NSs in a new aspect.

\bigskip
The authors thank to Ken'ichiro Nakazato, Matthias Hempel, Shun Furusawa, Nihal Buyukcizmeci,
Igor Mishustin and Tsuyoshi Miyatsu for kindly providing data of their theoretical models.
T. S. and C. I. thank to Masayuki Y. Fujimoto
for fruitful discussions on neutron star mass and radius,
and to Jirina R. Stone, Thomas Klaehn and Micaela Oestel for useful discussions on the database.
CI has been supported by Sasagawa Grants for Scientific Fellows of the Japan Science Society, 
F12-208.
This work has been partially supported by Grant-in-Aid for Scientific Research on Innovative Areas (20105001, 24105001)
and by Grant-in-Aid for Scientific Research (22540296, 23224004, 24244036, 26105515, 26104006), from Japan Society of the Promotion of Science.
The authors appreciate the support of Tokyo University of Science
and Research Center for Nuclear Physics, Osaka University
for the use of the database server.

\bibliographystyle{apj}
\bibliography{apj-jour,reference}

\end{document}